\documentclass[runningheads]{llncs}
\usepackage{graphicx}
\usepackage{tikz}
\newcommand*\circled[1]{\tikz[baseline=(char.base)]{
            \node[shape=circle,draw,inner sep=0.5pt] (char) {#1};}}

\begin{document}
\title{ARTeX: Anonymity Real-world-assets Token eXchange}
%
%
\author{Jaeseong Lee\orcidID{0009-0001-7582-9927} \and
{Junghee Lee\orcidID{0000-0003-0733-0136}
}} 
\authorrunning{Jaeseong Lee et al.}
%
\institute{Korea University, School of Cybersecurity \\
145, Anam-ro, Seongbuk-gu, Seoul, Republic of Korea\\
\email{\{{mandoopapa, j\_lee}\}@korea.ac.kr}}
\maketitle              
\begin{abstract}
This paper addresses one of the most noteworthy issues in the recent virtual asset market, the privacy concerns related to token transactions of Real-World Assets tokens, known as RWA tokens. Following the advent of Bitcoin, the virtual asset market has experienced explosive growth, spawning movements to link real-world assets with virtual assets. However, due to the transparency principle of blockchain technology, the anonymity of traders cannot be guaranteed. In the existing blockchain environment, there have been instances of protecting the privacy of fungible tokens (FTs) using mixer services. Moreover, numerous studies have been conducted to secure the privacy of non-fungible tokens (NFTs). However, due to the unique characteristics of RWA tokens and the limitations of each study, it has been challenging to achieve the goal of anonymity protection effectively. This paper proposes a new token trading platform, the ARTeX, designed to resolve these issues. This platform not only addresses the shortcomings of existing methods but also ensures the anonymity of traders while enhancing safeguards against illegal activities.

\end{abstract}
%
%
\section{Introduction}

Since the introduction of Bitcoin by the enigmatic figure, Satoshi Nakamoto, in 2008, numerous alternative coins have been created, fostering the growth of a new industry \cite{nakamoto2008bitcoin}. As the digital asset market thrives, initiatives to link traditional physical assets with digital assets are emerging due to the advancement of these technologies. The RWA token is an acronym for Real-World Assets token, which symbolizes the connection with real assets. The RWA token is considered a concept that encompasses not only STO (Security Token Offering) \cite{lambert2021sto}  but also non-exchangeable NFT (Non-Fungible Token), and SBT (Soulbound Token)~\cite{chaffer2022SBT}. 

Due to the technological characteristics of the blockchain, transaction histories and information are transparently disclosed, which is also the case with RWA tokens. Ethereum, a representative public blockchain, allows you to easily inspect specific token transaction histories and detailed information through third-party applications like Etherscan \cite{oliva2022mining}. At this point, privacy issues related to token transactions are raised. Many details related to the token, such as who traded when, how, and at what price, are transparently disclosed and can be easily checked. Especially for traders, the trader's wallet address is provided in plain text. Through Etherscan, one can verify a wallet address presented in plaintext; not only the status of the tokens held by that wallet address, but also the transaction history can be ascertained. It is facile to determine who has transferred which tokens to whom. 

Efforts have been ongoing to solve this in various ways, as there is no proper device to hide transaction details related to their tokens and protect the anonymity of the traders. However, given that many of the proposed technologies to date have been confined to the protection of anonymity, there is a need for continued research into measures that encompass the protection of anonymity, including the actual service process. Therefore, in this paper, to overcome these problems, we examine the previously proposed anonymity protection methods, analyze the problems that occur in each method, and propose ARTeX, a token trading platform designed to protect anonymity.

\section{What is RWA Token?}

The Real-World Assets token has become widely known to the public by 2023, but discussions about the RWA token have already existed since 2017 \cite{notheisen2017trading}. Research has been conducted on how to trade real-world assets on the blockchain. The Real-World Assets token refers to the tokenization of all tangible and intangible assets existing in reality, and the RWA token defined in this paper is the same. It can also be seen as a concept encompassing the commonly mentioned NFT. Still, if NFT is a concept that encompasses simply non-exchangeable tokens, RWA can be seen as a concept that encompasses everything that can be tokenized, including assets linked to the real world. The proposed ERC3643 protocol for this RWA token standardization defined the concept of RWA tokens as including four things: real assets, securities, cryptocurrencies, and royalty programs \cite{erc3643}. According to this ERC3643 standard, all contracts traded following this standard are designed to be traceable. This is called the Identity Registry Contract, and the RWA tokens issued under ERC3643 support a contract that tracks the ownership of the token from the time of design by the issuer and the user \cite{erc3643}. 

\section{Anonymity Protection}
\subsection{Privacy Protection vs Anonymity}
Before we delve into the ARTeX platform, and the service we propose, we first explain why we emphasize anonymity among the two terms, anonymity, and privacy protection. Many users are using these two terms interchangeably. The reason we distinguish them is that ARTeX aims to prevent only the personal information of the sellers and buyers of the information from being revealed, not the token information itself. According to the terminological distinction between anonymity and privacy protection explained by Bradbury \cite{BRADBURY201410}, ARTeX is closer to hiding the owner of the secret rather than hiding the secret itself.

Within the framework of blockchain systems, anonymity can be viewed as a safeguard against the exposure of transactional details, encompassing the identities of both the sender and receiver, to all nodes. In a similar vein, privacy protection can be construed as the act of concealing the specifics of a transaction, such as the financial value involved. When a platform that places a premium on and safeguards anonymity is established, it inherently fortifies the privacy of its participants, subsequently offering a benefit akin to personal data protection from the perspective of the end-user. In its endeavor to maintain this degree of anonymity, ARTeX has been meticulously architected to ensure untraceability. This is achieved by instituting a distinct separation between the sellers and buyers of real-world assets, promoting an environment of unlinkability, and securely conveying transaction results (tokens) via a secure channel.

\subsection{Why Anonymity Protection is Necessary in RWA Token Transactions}
The anonymity of the buyer's wallet address, the seller's wallet address, and the transaction amount can be protected through obfuscation methods. Consider, for example, the environment of a blind auction for artwork. Let's assume that only anonymous bid proposals are received in an environment where the identity of who bid is not disclosed. Here, the artwork should be made public so that everyone can confirm what the piece is, ensuring that the purpose of the auction they are participating in is guaranteed even in an environment where they do not know who else is bidding. If some information about the artwork is also kept secret here, it would be difficult to be confident about the auction item they are bidding on, and there might be a feeling of anxiety that someone else, who is competing in the bid, might be attempting to deliberately pump the price, which could lower the overall trust in the auction. Not disclosing the identity of the seller here is not a common practice in art auctions, but in the transaction of the RWA tokens proposed in this paper, if the seller's identity, that is, the seller's wallet address, is exposed, there is a risk that the buyer's information can also be easily exposed. 
The platform should be one where the artwork itself can be assigned a fair value, and in that sense, it will be able to fully maintain its credibility. Given the traceability, it would be desirable to promote transactions of high-priced artworks that are hesitant because of this, by also hiding the transaction amount to facilitate the transaction process. 
If accurate information about the token is not provided to potential buyers, there is a risk of trading the wrong token or a token different from what they expected. It would naturally be necessary to support buyers in confirming whether the token they wish to purchase is accurate, to base the transaction on trust. 

\section{Related Works} 
\subsection{Hiding at the Frontend}
In a centralized marketplace, all information about the RWA token is disclosed, but the identities of the seller and buyer are hidden in the frontend, which is the most common method for ensuring anonymity in a typical web environment. However, as mentioned earlier, tokens on the blockchain have Explorer services that allow you to see the transaction information of the traded tokens at a glance, and you can immediately check the wallet addresses of the seller and buyer just by confirming the information of the traded tokens.

\subsection{Hiding Token Details} 
To overcome the shortcomings of the general method, a method of hiding the trading account of the seller and buyer within a centralized marketplace has also been proposed \cite{galal2022aegis}. In Aegis, they proposed a method of encrypting information on transaction amounts and target tokens (in this case, limited to NFTs) \cite{galal2022aegis}.
However, using this method in RWA token transactions is risky. For RWA token transactions, it is necessary to at least disclose detailed information about the tokens, as RWA tokens are based on real-world assets, and it is necessary to clearly define what assets they are based on. This corresponds to the point that the information necessary for the buyer (investor) must be provided.

\subsection{Decoy Accounts} 
A decoy account is a large-scale account managed by a centralized marketplace and is seen as equivalent to a typical EOA (Externally Owned Account) from the perspective of other users. However, the private key of the decoy account is managed by the marketplace and used to make it difficult to connect buyers and sellers in the NFT transaction process by separating cryptocurrency payments from NFT transfers \cite{chen2022toward}. When trading NFTs, the buyer pays the purchase cost to a decoy account, and the marketplace transfers the same amount that the buyer purchased from another decoy account to the seller \cite{chen2022toward}. Due to this process, it becomes impossible for others to find the link between the buyer, seller, and transferred NFT. This payment method via cryptocurrency appears as two separate general blockchain transactions between EOAs.

The method of using a decoy account is a good way to break the link between the seller and buyer, but it is difficult to have actual usability due to gas fees in the blockchain network. If only a minimal number of transactions occur within the platform for a certain period, anonymity can be enhanced through a decoy account, but if many transactions occur, many decoy accounts must be operated \cite{chen2022toward}, and each RWA token must hold a certain amount or more of native tokens in the blockchain where it is registered for gas fee utilization. This gas fee can be spent as a kind of security cost used to enhance anonymity, but if the transaction amount is not large, a situation where the cure is worse than the disease can occur.

The various ways proposed so far have suggested ways to enhance anonymity in token transactions, but with the attention on RWA tokens, an environment has been created that requires a focus on this. ARTeX proposes a new methodology that can enhance anonymity while overcoming the above-mentioned problems.

\section{Introduction to ARTeX}
\subsection{Overview of ARTeX}
In the case of traditional peer-to-peer transactions, there is a difficulty that the seller must directly find a potential buyer for their token. Even if a buyer is found with difficulty, it is hard to guarantee trust in mutual transactions. Even if they agree to deliver their tokens first and receive payment later, it can be hard to be certain that they can receive the payment due to the thoroughly non-face-to-face, anonymous transaction. The general trading market aims to bring sellers and buyers together in one place and safely resolve trust issues between them. We assume a virtual market called ARTeX to effectively represent the transaction process to achieve the anonymity of RWA token traders, and we introduce how to enhance anonymity by following the market process. ARTeX provides an open space on the web server for trading RWA tokens through listing at the seller's free will of the RWA token. It uses a familiar social market user interface (UI) so that users can list for token sales in a familiar UI environment, obtain information about tokens, have purchase intentions, and trade.

\begin{figure}[hbt!]
    \centering
    \includegraphics[width=130px]{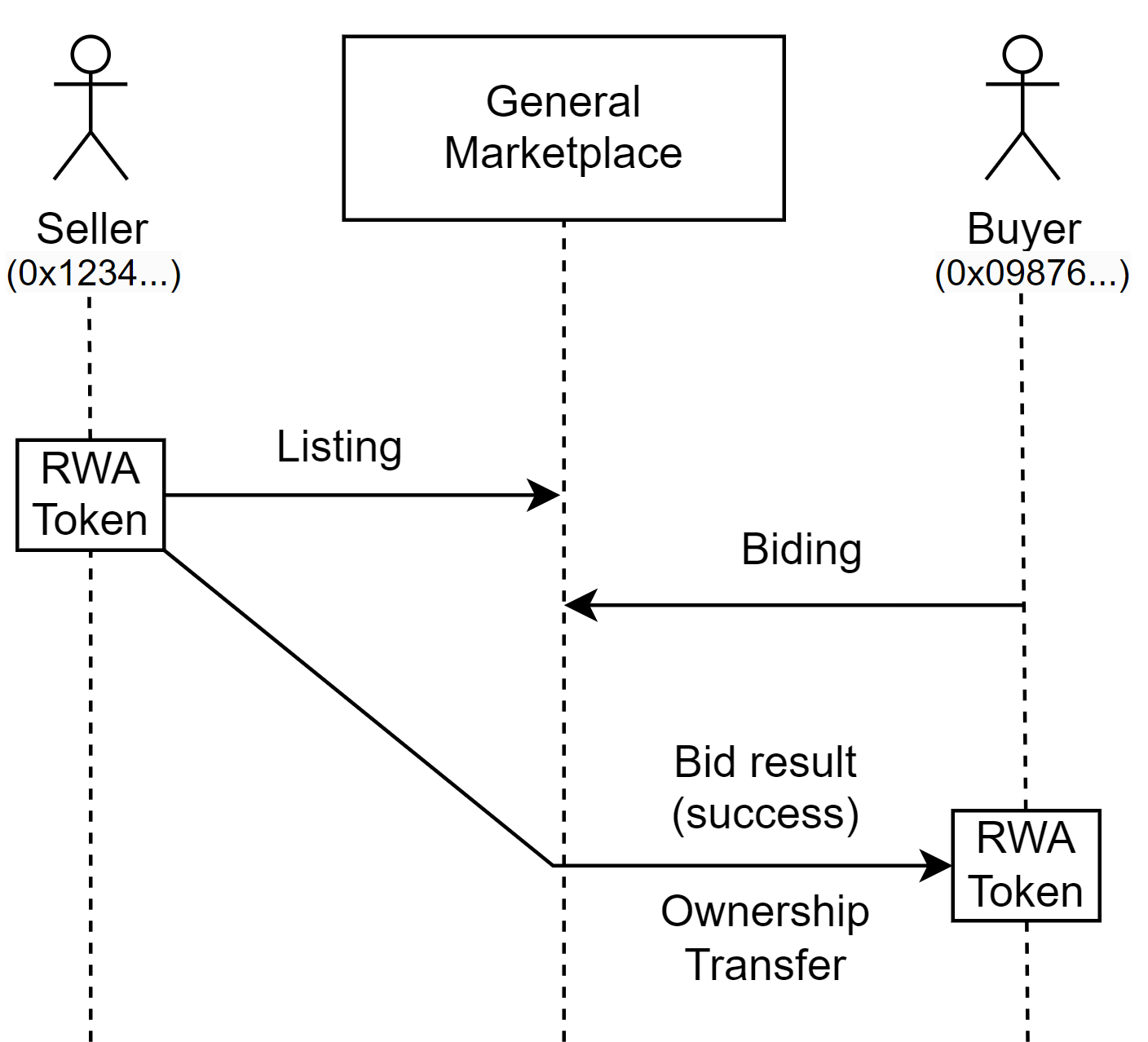}
    \caption{The case of trading RWA tokens in general}
    \label{fig:RWA token trade in general}
\end{figure}

\subsubsection{Trading RWA Tokens in general}
Let us first look at the case of trading RWA tokens in a typical market with Figure~\ref{fig:RWA token trade in general}. First, the seller lists the information of the token they want to sell on the market, exposing it to potential buyers. Buyers decide to purchase after viewing the listed tokens, and bid the appropriate amount. Depending on the situation, the seller may set up a sales contract at the appropriate price, or they may sell the token at the price of the winning bid after receiving the highest bid for a certain period. The sales process provides various options in the market to allow sellers and buyers to choose, and some markets only have one choice to take differentiation as a marketing strategy.

ARTeX goes further here, focusing on protecting the anonymity of sellers and buyers. As mentioned earlier, while playing the role of a platform that brings together sellers and buyers to trade in one place, it helps to get clear information about what token to buy while not wanting to reveal their identity due to the nature of the transaction. ARTeX introduces a process that differs from transactions in existing markets to enhance trust in the token trading market.

\subsubsection{Trading RWA Tokens on ARTeX}

\begin{figure}[hbt!]
    \centering
    \includegraphics[width=200px]{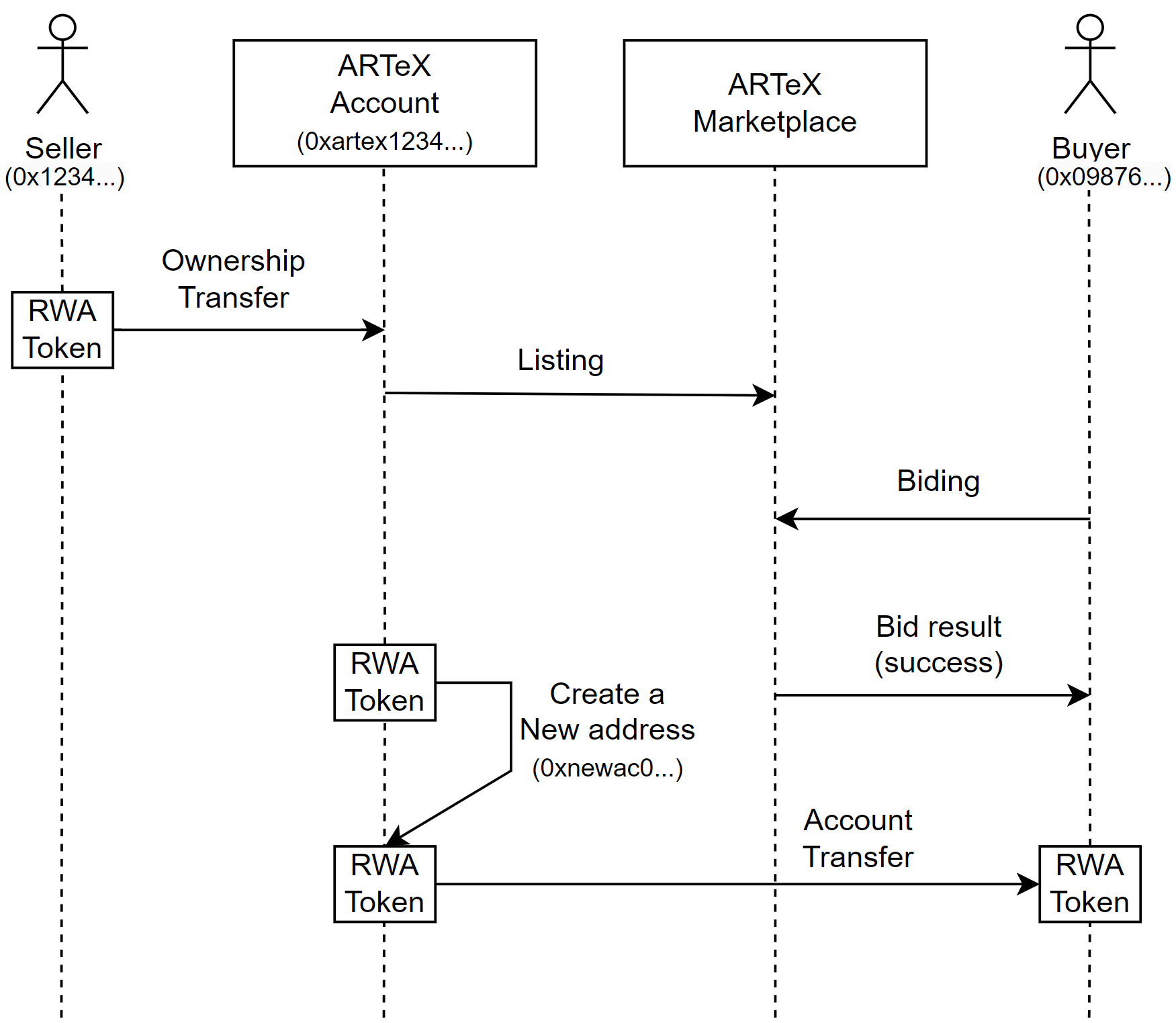}
    \caption{The case of Trading RWA Tokens on ARTeX}
    \label{fig:RWA token trade in ARTeX}
\end{figure}

 The seller sends RWA tokens to ARTeX for sale registration. The received RWA tokens are listed on ARTeX, allowing buyers to check the RWA tokens and make decisions for purchase. The information that can be publicly disclosed to potential buyers regarding the listed RWA tokens are as follows. Sellers can choose to disclose this information if necessary, and there is a need to provide at least minimal information for potential buyers. Within Figure~\ref{fig:RWA token trade in ARTeX}, individual wallet addresses are displayed to clearly show the difference between the Listing process and the Packaging process for sending to bidders. However, in reality, these wallet addresses are not publicly exposed, so no one can be verified by anyone.

Information that should be provided includes:
\begin{itemize}
    \item Token Contract Address: An address containing basic distribution information of the token, which can confirm accurate information of the token, such as metadata information.
    \item Token ID: If a token is non-fungible, it can be distinguished through the token ID.
    \item Token Standard: Confirm which standard the token to be sold was Minted according to.
    \item Token Amount: Indicates the number of tokens to be sold. In the case of tokens issued as Non-Fungible, the Token Amount becomes 1.
\end{itemize}

Information that can be provided if the seller wants:
\begin{itemize}
    \item Token Info: The seller can write a description to introduce the information of the token. It is recommended to write to help buyers' purchase decisions.
    \item Creator: The seller can disclose the user who first minted the token. It can be in the form of a wallet address, but it can also be set as a pseudonym.
    \item Image URL: If there is a separate image for the metadata included in the token or the description of the token, the seller can disclose it.
\end{itemize}


\subsubsection{Membership Registration and KYC Before Transactions}

Prior to transactions, both sellers and buyers must register on ARTeX. ARTeX collects and stores Know Your Customer (KYC) information necessary for settlement during sales and purchases \cite{george2019kyc}. KYC information may include passports, government-issued IDs or driver's licenses, social security numbers, etc., and documents that can prove that the registered member is the party to the transaction are also required. KYC information is necessary to prevent fraud, money laundering, and other illegal activities, and is necessary for investigation cooperation when tokens issued for illegal purposes are identified. Any information related to the identity of the seller and buyer will not be disclosed to the public, nor will it be shared even between the parties of the transaction.

After writing the ID, password, and currently used email through the ARTeX webpage and submitting it, the user will complete the provisional membership registration, and the user can browse ARTeX by logging in with the ID and password, but transactions cannot be made yet. Here, after entering and submitting the KYC information according to the field where the user enters the KYC information, the user will be registered as a full member who can trade RWA tokens once the review is completed. The following is a comprehensive summary of the information needed for ARTeX membership registration, which is a preparation process for transactions, and ARTeX keeps this information safe and does not share it with the outside or between the parties to the transaction \cite{conklin2004password}.
At ARTeX, as with general e-commerce, the procedures for selling, buying, and settling tokens may not occur consecutively. Therefore, various web service technologies that can confirm the credentials of ARTeX members, such as cookies or session ID, may be needed in the login process.

\subsection{ARTeX Trading Process}

\begin{figure}[hbt!]
    \centering
    \includegraphics[width=250px]{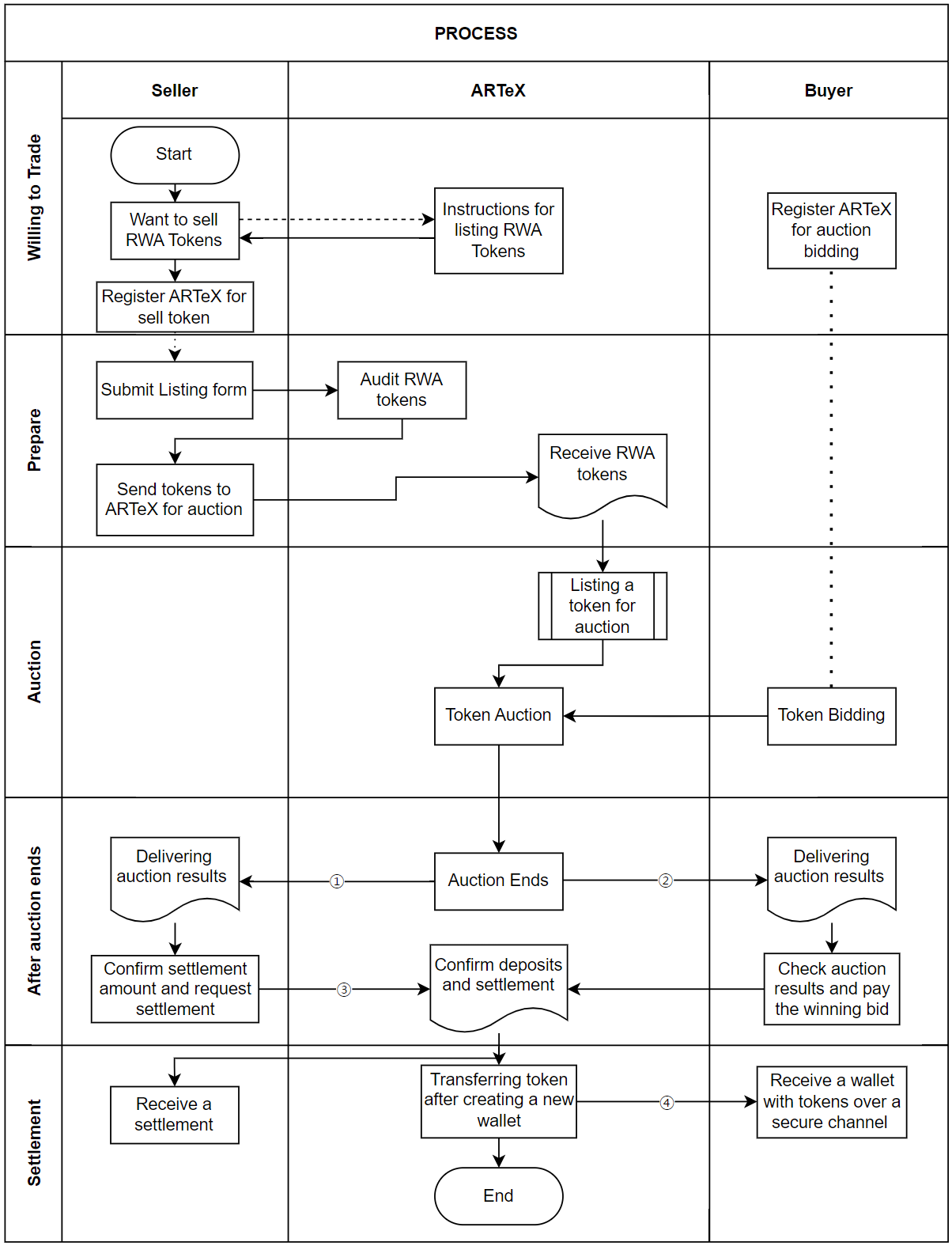}
    \caption{ARTeX Trading Process}
    \label{fig:RWA token trade}
\end{figure}

The ARTeX trading process is summarized in Figure~\ref{fig:RWA token trade}, which consists of five stages.
Each stage is described in detail in the following paragraphs.

\subsubsection{Token Holder Willing to Trade}
This is the stage where the token holders wish to sell their tokens. No token movement occurs at this stage. They confirm how to list their RWA tokens on the ARTeX market and, following the guide, decide to sell their tokens. At this stage, they check the guide on what information needs to be disclosed for the listing of RWA tokens on ARTeX and register on ARTeX. The seller, who has created an account $ID_{s}$, enters the information necessary for auction registration for the sale of RWA tokens and prepares for token sales.

\subsubsection{Prepare Auction}
The seller transfers the RWA token, $T_{rwa}$, to ARTeX for sale. Once the token transfer is confirmed, ARTeX enters the review before selling the token. ARTeX meticulously examines whether the token is issued properly and whether there is any illegality. Once it is confirmed that the token is properly issued and there is no problem, the token undergoes a kind of commodification process, gets listed on the ARTeX website, and completes all preparations for the auction.

\subsubsection{Auction}
Users wishing to purchase registered RWA tokens meticulously inspect the token information through ARTeX and bid in the auction. Once the auction finally concludes, the auction results are delivered to the token seller and the winning bidder. Apart from the final successful bidder, it is announced that the token auction has ended without any separate notice. Users wishing to purchase create their own $ID_{B}$ account that can be used within ARTeX to participate in the bidding. If ${ID_B}_1$ proposes 1 ETH, ${ID_B}_2$ can make a higher bid such as 1.1 ETH, and the auction proceeds in this way. According to game theory \cite{laffont1997game}, auctions are divided into the real-time public auction type, English auction, Dutch auction, and the private progress auction type, etc. Which method to adopt in ARTeX can be an important system and marketing element in actual service, but it is irrelevant to the main point of this paper, anonymity, so it is assumed that the auction was conducted safely in an arbitrary manner. However, there is research \cite{vickreyauction} suggesting that not disclosing bid price information to auction participants ensures transactions at fair prices in auctions. Therefore, even if the technical environment is possible, it would be advisable to conduct the auction in a direction that does not disclose bid price information to auction participants.

\subsubsection{After Auction Ends}
This is the stage where the auction concludes after a certain period. When the auction is over, ARTeX \circled{1} informs the seller that the auction has ended and what the final winning bid is. \circled{2} It confirms to the final successful bidder that the auction has ended and that they are the final successful bidder, simultaneously providing options for the winning bid amount and deposit method. The buyer can transfer the winning bid amount from one wallet to ARTeX's account all at once, or can pay in installments from multiple accounts. Depending on the option, they can confirm as many ARTeX wallet addresses as they want. In other words, the final successful bidder can confirm the ARTeX wallet address to be paid according to the auction end fact, confirmation that they are the final successful bidder, the winning bid amount, and the deposit method option.
\circled{3} The token seller confirms the final winning bid and requests ARTeX for settlement. The seller conveys that they have agreed to the settlement amount to be auctioned, and at the same time, provides the wallet address to receive the settlement amount to ARTeX. At this time, the seller can receive the settlement money to the wallet address ${{T_{sell}}_1}$ where they owned the token, and as another option, they can receive the settlement money to another wallet address ${{P_{sell}}_2}$ they own. Generally, the latter case will be slightly more advantageous for anonymity protection. If the seller receives the settlement amount to another wallet address, it becomes very difficult to confirm how much the seller has settled for delivering the token to ARTeX when checking the details of the sold RWA token.
The token buyer delivers the final bid amount in the auction to ARTeX. The settlement wallet address that receives this is ${{P_{sell}}_2}$, a separate wallet from the wallet ${{P_A}_1}$ received RWA token from the token seller. They can transfer the entire amount from one wallet address, or they can pay in installments. For example, if it was auctioned at 100 ETH and this is transferred, it is possible to ${{P_{buy}}_1}$ (100 ETH) → ${{P_A}_2}$ or ${{P_{buy}}_1}$(50 ETH) + ${{P_{buy}}_2}$(30 ETH) + ${{P_{buy}}_3}$(20 ETH) → ${{P_A}_2}$.
Similarly, the wallet address that receives the transfer to ARTeX does not necessarily have to be one of ${{P_A}_2}$, and the form of ${{P_{buy}}_1}$(90 ETH) → ${{P_A}_2}$ and ${{P_{buy}}_2}$(10 ETH)→ ${{P_A}_3}$ is also possible. If the total amount of ARTeX is delivered, there will be no problem with the final settlement.
When it is confirmed that the settlement amount has been properly deposited, ARTeX creates a separate wallet to deliver to the token buyer and delivers the sold RWA token to the corresponding wallet address.

\subsubsection{Settlement}
The seller of the token receives the sales settlement money to the wallet address ${{T_{sell}}_1}$ or ${{T_{sell}}_2}$ where they want to receive the settlement money, and the sales process is completed. \circled{4} The new wallet ${{T_A}_{new}}$ containing the RWA token is delivered to the buyer along with the private key through the secure channel, and the purchase process is also completed. ARTeX announces that the transaction of the token is completely finished.

\subsection{Secure Channel for Delivering the Wallet Containing the Token}
The method of delivery through a separate secure channel is not commonly used in existing blockchain-based transaction businesses like NFT marketplaces. However, to protect the anonymity of RWA tokens, the use of a secure channel is considered and applied to the ARTeX model. ARTeX receives tokens from sellers for completed transactions, creates a new wallet address, and delivers it to the buyer through a separate secure channel. By using this method, buyers can securely trade the desired RWA tokens without exposing their wallet addresses and complete transactions. After purchase, the buyer can freely transfer the token to a personal wallet as needed, and if an investigation is required due to the suspected illegality of the traded token, ARTeX can provide relevant information because it has the wallet address delivered to the buyer through the secure channel when an investigation cooperation request is received from the investigator. Subsequent tracking will be possible up to the steps recorded on-chain. ARTeX can provide it according to the legal procedures, and even if it does not provide the communication results through a secure channel, tracking is possible as long as the investigative agency knows the exact information of the token.
The safety of the secure channel itself is a security issue that all centralized commerce services are equally experiencing, so it will not be dealt with separately in this paper.

\section{Analysis}
If the process of ARTeX is followed, we analyze how the linkability between RWA token sellers and buyers can be severed to achieve anonymity.

\begin{figure}[hbt!]
    \centering
    \includegraphics[width=170px]{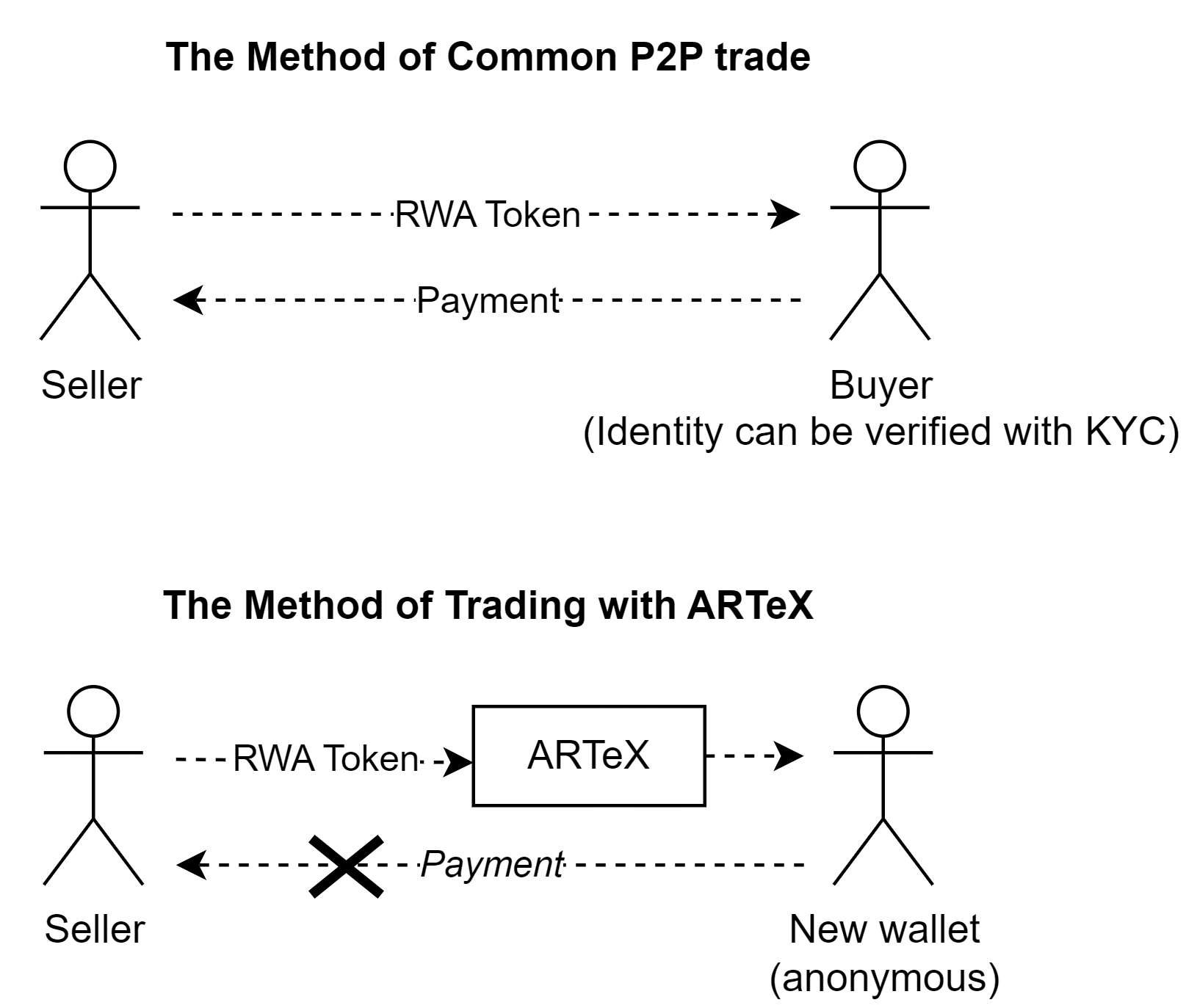}
    \caption{Difference between Common P2P Trade and ARTeX Trade}
    \label{fig:Difference between Common P2P Trade and ARTeX Trade}
\end{figure}

In a common peer-to-peer trade, both the Seller and the Buyer execute an exchange where a certain amount is paid and equivalent tokens are received. This is a routine transaction, and through the process of KYC, the identity of the buyer can be ascertained. However, in transactions conducted via ARTeX, while RWA tokens are indeed transferred, the equivalent monetary amount is not directly received. Under such circumstances, a third party may find it challenging to identify it as a transaction based solely on the token information. It could be perceived as a donation, a gift, or a consignment, or even a misappropriation due to fraudulent activities. It becomes notably difficult to discern for how much and to whom the token was traded. Although the token has been transferred, the anonymity of the transaction remains intact. This method is feasible because it severs the link between the transaction parties.

\subsection{Breaking the Link between Token Sellers and Buyers}
The primary ingredient in preserving anonymity is to successfully blur the connection between the token seller and the buyer. Simultaneously, severing the transaction link, which renders the transaction unnoticeable when observed from an external viewpoint, constitutes a pivotal strategy for upholding privacy.
In conventional transactions, the RWA token is transferred from the vendor's wallet address ${{T_{sell}}}$ to the purchaser's wallet address ${{T_{buy}}}$. Additionally, the transaction payment is also transferred from ${{P_{sell}}}$ to ${{P_{buy}}}$. This transaction record is preserved on the blockchain. Should parties engage in direct P2P transactions as outlined above, the public can easily ascertain the details of the transaction: the token involved, the timing, the amount transacted, the identities of the vendor and purchaser. This information can be obtained by merely verifying the token's transaction. As these records are perpetually stored on the blockchain, even as time progresses and more transactions occur, it remains straightforward to review past records. The identities of the participants and the token's price from a year ago or even ten years ago can be readily accessed. The inherent nature of the blockchain, which renders data alteration and deletion progressively more challenging over time, further compounds the difficulty of achieving anonymity if the transaction link is not disassociated initially.

The transaction form of tokens, following the ARTeX process, is transferred in the order of ${{T_{sell}}_1}$ → ${{T_A}_1}$→ ${{T_A}_{new}}$. The delivery of the transaction amount can take various forms: \\
(1) The buyer pays the total bid amount at once, and the seller receives the settlement in the wallet P that held the token: 
$${{P_{buy}}_1} \to {{P_A}_2} \to {{P_{sell}}_1}$$
(2) The buyer pays the total bid amount at once, and the seller receives the settlement in a different wallet, not the wallet that held the token: 
$${{P_{buy}}_1} \to {{P_A}_2} \to {{P_{sell}}_2}$$
(3) The buyer divides the bid amount into multiple wallets and pays, and the seller receives the settlement in the wallet that held the token: 
$${{P_{buy}}_1} + {{P_{buy}}_2} + {{P_{buy}}_3} \to {{P_A}_2} \to {{P_{Sell}}_1}$$
(4) The buyer divides the bid amount into multiple wallets and pays, and the seller receives the settlement in a different wallet, not the wallet that held the token: 
$${{P_{buy}}_1} + {{P_{buy}}_2} + {{P_{buy}}_3} \to {{P_A}_2} \to {{P_{Sell}}_2}$$
(5) Depending on the case, the seller can also divide the settlement into multiple wallets: 
$${{P_{buy}}_1} + {{P_{buy}}_2} + {{P_{buy}}_3} \to {{P_A}_2} \to {{P_{sell}}_2} + {{P_{sell}}_3} + {{P_{sell}}_4}$$ 
(6) It is also possible to divide the bid amount into multiple wallets and deliver it in ARTeX. Specifically, this can be seen as the strongest in protecting anonymity: 
$${{P_{buy}}_1} + {{P_{buy}}_2} + {{P_{buy}}_3} \to {{P_A}_2} + {{P_A}_3} + {{P_A}_4} \to {{P_{sell}}_2} + {{P_{sell}}_3} + {{P_{sell}}_4}$$ 

In here, ${{P_{buy}}_1}$ does not necessarily have to match ${{P_A}_2}$ and ${{P_{sell}}_2}$, and they can mix and trade. Also, the settlement does not have to occur at the same time. In this case, if you agree with the party receiving the settlement, it is also possible to receive a little bit of settlement over several days in a total of n times. In this case, it becomes more difficult to confirm the exact transaction amount.

If the process of ARTeX is followed, the token transaction will appear as follows when confirmed. If the seller's wallet that held the token is the standard, the token is transferred from ${{T_{sell}}_1}$ to ${{T_A}_1}$, and no separate amount is received. The wallet ${{T_A}_1}$ that received the token from ${{T_{sell}}_1}$ transferred the token to ${{T_A}_{new}}$, but did not receive a separate amount. Here, the token transfer ends. There was no payment for this. Therefore, if you check this transaction, you can confirm that the token was transferred from someone to ${{T_A}_1}$ and then to ${{T_A}_{new}}$, but since the ${{T_A}_{new}}$ wallet is a newly created wallet, there is no previous transaction history, so you cannot confirm who it is. In other words, it is difficult to confirm who bought the token or whether this transaction is a transaction. It could be a donation, not a transaction, or a gift between acquaintances. Seriously, it could have been a phishing scam, or it could have been sent by mistake due to a typo in the address. It is difficult to claim that it is a mutual transaction just by looking at the above transaction form.

\subsection{The reason for not using mixer services for RWA token trading}
In the case of tokens of commonly used blockchain networks, it is possible to mix and hide transaction details using Mixer services. RWA tokens can be issued in a Non-fungible form depending on the case, and since they are usually issued linked to physical assets and the amount of issuance is fixed, the possibility of achieving their purpose by using Mixer is markedly low. Therefore, in the case of transactions such as RWA tokens, it would be reasonable to protect anonymity by trading through a trustworthy third party like ARTeX. And in the case of Mixer, there is a controversy that it can be deeply related to illegality. Binance blocked withdrawals to Wasabi, a privacy-protecting Bitcoin wallet that integrated the famous mixer service CoinJoin, in 2019 \cite{binanceWasabi}. Binance announced at the time that there were circumstances which funds related to money laundering had entered through the CoinJoin service. 

\section{Conclusion}
In high-value or art token transactions, the set of processes proposed by ARTeX helps both the owner who wants to sell and the collector who wants to buy to conduct transactions without explicitly revealing their personal information on the blockchain. This paper is not merely a service proposal but presents how a systematized transaction process established in the real world can be applied to Web3. The few technologies used for personal information protection included here and the process that enables the organic combination of these technologies will serve as a foundation for making a practical service possible by shaping its form more clearly through future research. However, it is difficult to propose a protocol that completely disconnects the information of the seller and the buyer. As mentioned earlier, losing the whereabouts of token transactions due to illegal tokens, hacking, fraud, etc. negatively affects the investigation by government authorities, which is also the case with the illegality of Mixer services. If a method that can verify safety and legality is proposed depending on the case, it would be possible to propose a process that can completely separate the connection between the seller and the buyer from the token on-chain.

\cite{gayialis2021approach}

%
%
%
%

\bibliography{ref}

\begin{thebibliography}{10}
\providecommand{\url}[1]{\texttt{#1}}
\providecommand{\urlprefix}{URL }
\providecommand{\doi}[1]{https://doi.org/#1}

\bibitem{BRADBURY201410}
Bradbury, D.: Anonymity and privacy: a guide for the perplexed. Network Security  \textbf{10},  10--14 (2014)

\bibitem{chaffer2022SBT}
Chaffer, T.J., Goldston, J.: On the existential basis of self-sovereign identity and soulbound tokens: An examination of the" self" in the age of web3. Journal of Strategic Innovation and Sustainability  \textbf{17}(3), ~1--9 (2022)

\bibitem{chen2022toward}
Chen, Z., Omote, K.: Toward achieving anonymous nft trading. IEEE Access  \textbf{10},  130166--130176 (2022)

\bibitem{conklin2004password}
Conklin, A., Dietrich, G., Walz, D.: Password-based authentication: a system perspective. In: 37th Annual Hawaii International Conference on System Sciences, 2004. Proceedings of the. pp. 10--pp. IEEE (2004)

\bibitem{galal2022aegis}
Galal, H.S., Youssef, A.M.: Aegis: Privacy-preserving market for non-fungible tokens. IEEE Transactions on Network Science and Engineering  \textbf{10}(1),  92--102 (2022)

\bibitem{gayialis2021approach}
Gayialis, S.P., Kechagias, E.P., Konstantakopoulos, G.D., Papadopoulos, G.A., Tatsiopoulos, I.P.: An approach for creating a blockchain platform for labeling and tracing wines and spirits. In: Advances in Production Management Systems. Artificial Intelligence for Sustainable and Resilient Production Systems: IFIP WG 5.7 International Conference, APMS 2021, Nantes, France, September 5--9, 2021, Proceedings, Part IV. pp. 81--89. Springer (2021)

\bibitem{george2019kyc}
George, D., Wani, A., Bhatia, A.: A blockchain based solution to know your customer (kyc) dilemma. In: 2019 IEEE International Conference on Advanced Networks and Telecommunications Systems (ANTS). pp.~1--6. IEEE (2019)

\bibitem{erc3643}
Krishna, G.C., IR, P.J.: Exploring the ethereum blockchain: An introduction to blockchain technology. In: Handbook of Research on Data Science and Cybersecurity Innovations in Industry 4.0 Technologies, pp. 261--290. IGI Global (2023)

\bibitem{laffont1997game}
Laffont, J.J.: Game theory and empirical economics: The case of auction data. European Economic Review  \textbf{41}(1),  1--35 (1997)

\bibitem{lambert2021sto}
Lambert, T., Liebau, D., Roosenboom, P.: Security token offerings. Small Business Economics pp. 1--27 (2021)

\bibitem{nakamoto2008bitcoin}
Nakamoto, S.: Bitcoin: A peer-to-peer electronic cash system. Decentralized business review  (2008)

\bibitem{binanceWasabi}
Nelson, D.: Binance blockade of wasabi wallet could point to a crypto crack-up. Coindesk  (2021), \url{https://www.coindesk.com/policy/2019/12/26/binance-blockade-of-wasabi-wallet-could-point-to-a-crypto-crack-up}

\bibitem{notheisen2017trading}
Notheisen, B., Cholewa, J.B., Shanmugam, A.P.: Trading real-world assets on blockchain: an application of trust-free transaction systems in the market for lemons. Business \& Information Systems Engineering  \textbf{59},  425--440 (2017)

\bibitem{oliva2022mining}
Oliva, G.A.: Mining the ethereum blockchain platform: best practices and pitfalls (msr 2022 tutorial). In: Proceedings of the 19th International Conference on Mining Software Repositories. pp. 201--202 (2022)

\bibitem{vickreyauction}
Vickrey, W.: Counterspeculation, auctions, and competitive sealed tenders. The Journal of finance  \textbf{16}(1),  8--37 (1961)

\end{thebibliography}
\bibliographystyle{splncs04}

\end{document}